\documentclass[twocolumn,secnumarabic,msmath,amssymb,floatfix]{revtex4}

\usepackage{graphicx}
\usepackage{subfigure} 
\usepackage{amssymb}
\usepackage{color}

\begin{document} \title{{\small {\it Chaos, Special Focus Issue on Nonlinear
Dynamics in Cognitive and Neural Systems.}}\\ Controlling the onset of
traveling pulses
in excitable media\\ by nonlocal spatial coupling and time-delayed feedback}

\author {Felix M. Schneider} \author {Eckehard Sch\"oll} \author {Markus A.
Dahlem}  \affiliation{Institut f{\"u}r Theoretische Physik, Technische
Universit{\"a}t Berlin,  Hardenbergstra{\ss}e 36, D-10623 Berlin, Germany}

\begin{abstract} 
  
The onset of pulse propagation is studied in a reaction-diffusion (RD) model
with control by augmented transmission capability that is provided either along
nonlocal spatial coupling or by time-delayed feedback. We show that traveling
pulses occur primarily as solutions to the RD equations while augmented
transmission changes excitability. For certain ranges of the parameter
settings, defined as weak susceptibility and moderate control, respectively,
the hybrid model can be mapped to the original RD model.  This results in an
effective change of RD parameters controlled by augmented transmission. Outside
moderate control parameter settings new patterns are obtained, for example
step-wise propagation due to delay-induced oscillations. Augmented transmission
constitutes a signaling system complementary to the classical RD mechanism of
pattern formation. Our hybrid model  combines the two major signalling systems
in the brain, namely volume transmission and synaptic transmission. Our results
provide insights into the spread and control of pathological pulses in the
brain.  

\end{abstract} 

 \pacs{} \maketitle

{\bf  Traveling pulses are of fundamental importance in neuroscience. They
propagate not only information along nerve fibers, but are also related to
pathological phenomena.  Examples are cell depolarizations that lead to a
temporary complete loss of normal cell functions in migraine and stroke.  This
state spreads in cortical tissue via chemical signals that diffuse through the
extracellular space.    We study these \textit{spreading depolarization} pulses
in a standard reaction-diffusion model and suggest to augment the transmission
capabilities such that they reflect the cortical structural and functional
connectivity. With this new modality, we investigate control of emerging spread
of pathological states in the brain.}

\section{Introduction}

Within the last years control of complex dynamics has evolved as one of the
central issues in applied nonlinear science \cite{SCH07}. Major progress has
been made in neuroscience, among other areas,  by extending methods of chaos
control, in particular time-delayed feedback \cite{PYR92}, to spatio-temporal
patterns \cite{DAH08,KEH09,KYR09}, and by developing applications in the field
of biomedical engineering \cite{RIC05,BAR08a}.  In this study, control is
introduced to suppress spatio-temporal pattern formation. Our emphasis is on
understanding the recruitment of cortical tissue into dysfunctional states by
traveling pulses of pathological activity, in particular, on  internal cortical
circuits that provide augmented transmission capabilities and that can
prevent such events.  Our long-term aim is to design strategies  that either
support the internal cortical control or mimic its behavior by external control
loops and translate these methods into applications.

There is growing experimental evidence that particular spatio-temporal pulse
patterns in the human cortex, called {\em cortical spreading depression},
cause transient neurological symptoms during migraine \cite{LAU94,HAD01}.
Similar pulses occur after stroke, called {\em periinfarct depolarization}, and
contribute to the loss of potentially salvageable tissue, i.\,e., tissue at
risk of infarction \cite{FAB06}.  These dysfunctional states of the cortex are
also referred to as {\em spreading depolarizations} (SD) to point out the
nearly complete depolarizations of cortical cells and its spread  as the
common aspect in these patterns.  

SD is usually called a cortical \textit{wave}, not pulse, which might cause some
confusion. In many cases these two terms can be used interchangeably.  We
adhere to a  precise mathematical terminology,  referring to a traveling pulse
as a localized event with a spatial profile having a single front and back
while a wave usually refers to a periodic spatial profile. SD is a pulse in
this terminology,  the strict use of which is necessary because the
spatio-temporal pattern formation of SD occurs in cortical tissue being only
weakly susceptible to SD \cite{DAH09}.  The definition of  \textit{weak
susceptibility} (Sec.~\ref{sec:lese}) depends on a bifurcation for which one
must strictly distinguish between  solitary and periodic wave forms. 

The pulse of SD extends in the cortex  over several centimeters with a
remarkably slow speed of several millimeter per minute.  Accordingly, the
mathematical description of SD considers large-scale neuronal activity  in
populations of neurons rather than ion channels in the membrane of nerve
fibers, and  the spatial coupling is provided by volume transmission, which is
essentially a diffusion process.  The first mathematical model of SD was
proposed by Hodgkin and Grafstein \cite{GRA63} based on bistable $K^+$
dynamics and $K^+$ diffusion. This model, however, describes only the front
dynamics of SD.  We suggest (Sec.~\ref{sec:rd}) two extensions to this model to
study the onset of SD.  The first is a necessary extension to obtain onset
behavior in the Hodgkin-Grafstein model.  We include in a generic form a
recovery process for the pulse trailing edge.  In a second step, we extend this
model further by nonlocal and time-delayed signal transmission to study the
effect of internal control provided as a feedback to the traveling pulse.
Results and conclusions are given in Sec.~\ref{sec:spcg} and
Sec.~\ref{sec:discussion}, respectively.

\section{Classifications of excitability in local elements and extended media}  

\label{sec:lese} 

Excitability, as a property of a single element, is based on threshold
behavior and therefore requires a nonlinear process with a stable fixed point.
If the system is sufficiently perturbed from this fixed point, it returns after
a large excursion in phase space, emitting a spike \cite{LIN04}. If excitable
elements are locally interconnected, a new behavior  can emerge, namely the
capacity to propagate a sustained pulse  through this spatially extended
system.  This emergent property  defines a medium as being excitable, also
termed an {\it active medium}.  Excitable elements and excitable media differ
in their response to superthreshold stimulation.  The response of an excitable
element to a superthreshold stimulation will eventually end in the stable fixed
point value of the steady state of this element. In contrast, the response of
an excitable medium, which is initially in the homogeneous steady state, to a
superthreshold stimulation results in approaching  a new attractor, the pulse
solution. This different behavior of local elements and spatial media indicates
that there are also different ways to classify excitability in elements and
media.

In Sec.~\ref{sec:le} definitions are provided for excitability  of local
elements with a focus on neural systems.  Some of our results concerning
control of spatial-temporal patterns can be explained by considering the
effect of control on the local dynamics (Sec.~\ref{sec:single_system}).
Furthermore, we will consider local dynamics with the aim to extend the
Hodgkin-Grafstein model (Sec.~\ref{sec:rd}).  However, our main focus is on
spatial excitability (Sec.~\ref{sec:se}) and its control by nonlocal and
time-delayed feedback described in Sec.~\ref{sec:nonlocal_control} and
Sec.~\ref{sec:TDAS}, respectively. For a thorough treatment of local dynamics,
in particular in neural systems, see for example Ref.~\cite{ERM98} or  for more
complex discharge patterns, like bursting, Ref.~\cite{IZH00}.

\subsection{Local excitability}

\label{sec:le} 

\begin{figure}[b] 
  \includegraphics[width=0.48\textwidth]{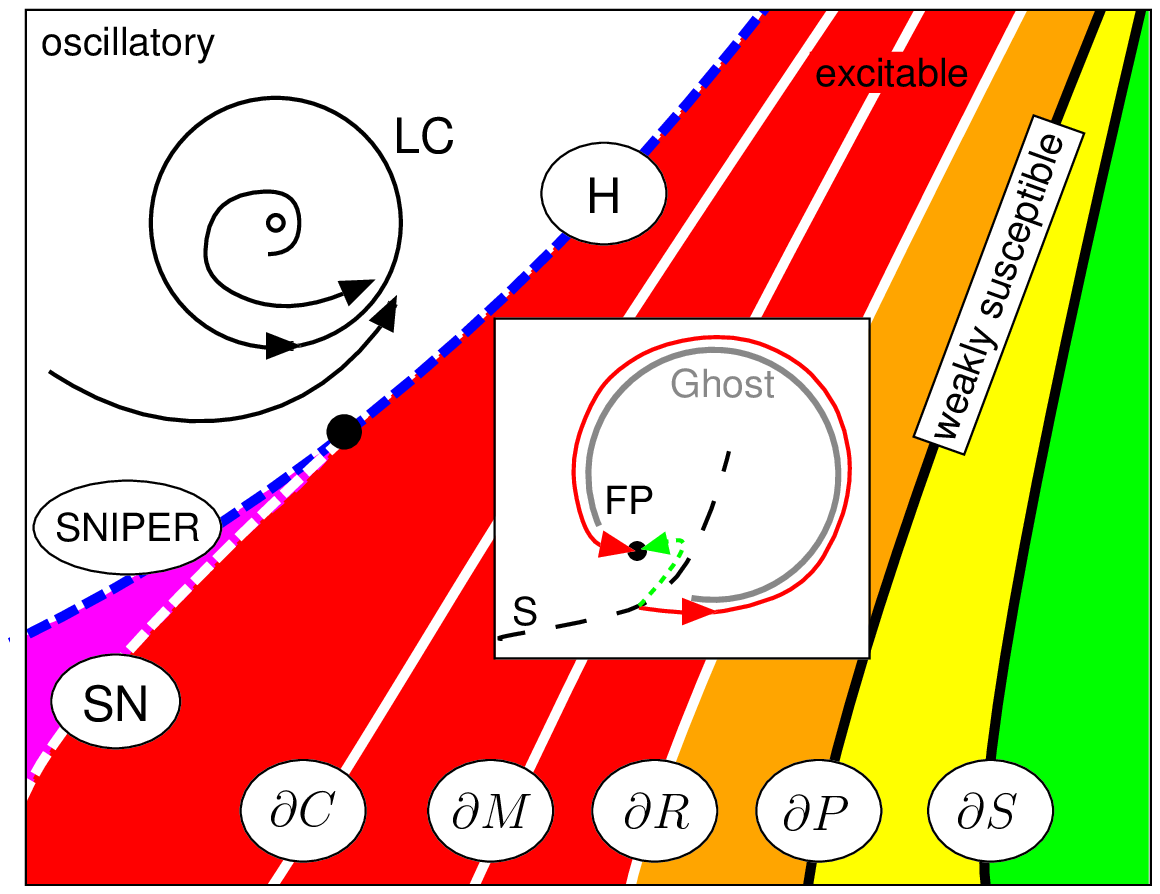}
  \caption{
  Parameter space of excitable systems illustrating a universal
  scheme for classifications of both local and spatial excitability.
  Bifurcation lines for excitable elements of an active media are marked by
  thick blue (dashed) lines. Transitions from the excitable (magenta, red,
  orange) to the oscillatory regime (white) occur usually either through a
  saddle-node  infinite period bifurcation (SNIPER) causing excitability of
  type I, or via a Hopf bifurcation (H) causing excitability of type II.
  Between the SNIPER and a further saddle-node (SN) bifurcation line (magenta)
  three fixed points (stable node, unstable focus, and saddle) exist in the
  local excitable elements.  The thick white solid lines $\partial C$,
  $\partial M$, and $\partial R$ mark bifurcations of active media where the
  spatio-temporal pattern in 2D changes (complex, meandering, and rigid
  rotating spiral patterns, to the left of $\partial C$, $\partial M$ and
  $\partial R$, respectively).  At $\partial P$ the medium becomes
  non-excitable.  Sustained pulses do not exist in the  yellow regime but
  there is a transient propagating pulse. To the right of $\partial S$
  (green) the transient activation radius becomes zero.  The insets (thin
  lines) indicate the corresponding configuration of trajectories in phase
  space, with limit cycle (LC), sharp separatrix (S), and fixed point (FP).} 
    {\label{fig:class}}
\end{figure}

A generic mechanism of local excitability  requires a certain configuration
of trajectories in the phase space of a single excitable element (inset of
Fig.~\ref{fig:class}). This configuration usually results from the parameter
vicinity of an oscillatory regime whose large amplitude limit cycle is suddenly
destructed \cite{IZH00}.  

The parameter space is schematically depicted in Fig.~\ref{fig:class}, where
the white area corresponds to the oscillatory regime, and the colored areas
mark various regimes of excitability and non-excitability.  If parameters
settings are in the excitable regime (to the right of the thick blue dashed line), a
rest state (fixed point) is the only attractor.
There exists also a threshold  close to this rest state, e.g. a "sharp"
separatrix or trajectory in phase space (thin dashed line, in the inset).
Trajectories starting on the near side of the threshold  (green dotted)
approach the rest state directly (subthreshold response), while those starting
on the far side (red solid) perform  a large excursion in phase space, guided
by the ghost of a  destructed large amplitude limit cycle, before returning to
the rest state (superthreshold response).  Note that we refer to excitable
\textit{elements} and not to neurons because also a population of neurons
described in a firing rate model can behave like an excitable element.

The way in which, by changing a bifurcation parameter, the dynamics of the
local element changes from excitable to oscillatory behavior, that is, the way
the large amplitude limit cycle is built up, is used to classify the type of
excitability in a single excitable element.  This can happen in different ways,
two of which are usually distinguished \cite{ERM98}.  Type I excitability
obtains its characteristic features from a saddle-node infinite period (SNIPER)
bifurcation:  the frequency of the emerging periodic orbit tends to zero,
while the amplitude starts with a finite value.  Type II excitability is caused
by a Hopf bifurcation, which is characterized by the features that the periodic
orbit emerges with zero amplitude and nonzero frequency. In practice, the
main characteristic for type II excitability is, however, the onset
frequency, because for this type a canard explosion renders the zero amplitude
practically invisible.

\subsection{Onset of pulse propagation in spatial excitability}
\label{sec:se} 

\begin{figure}[t]
  \includegraphics[width=0.4\textwidth]{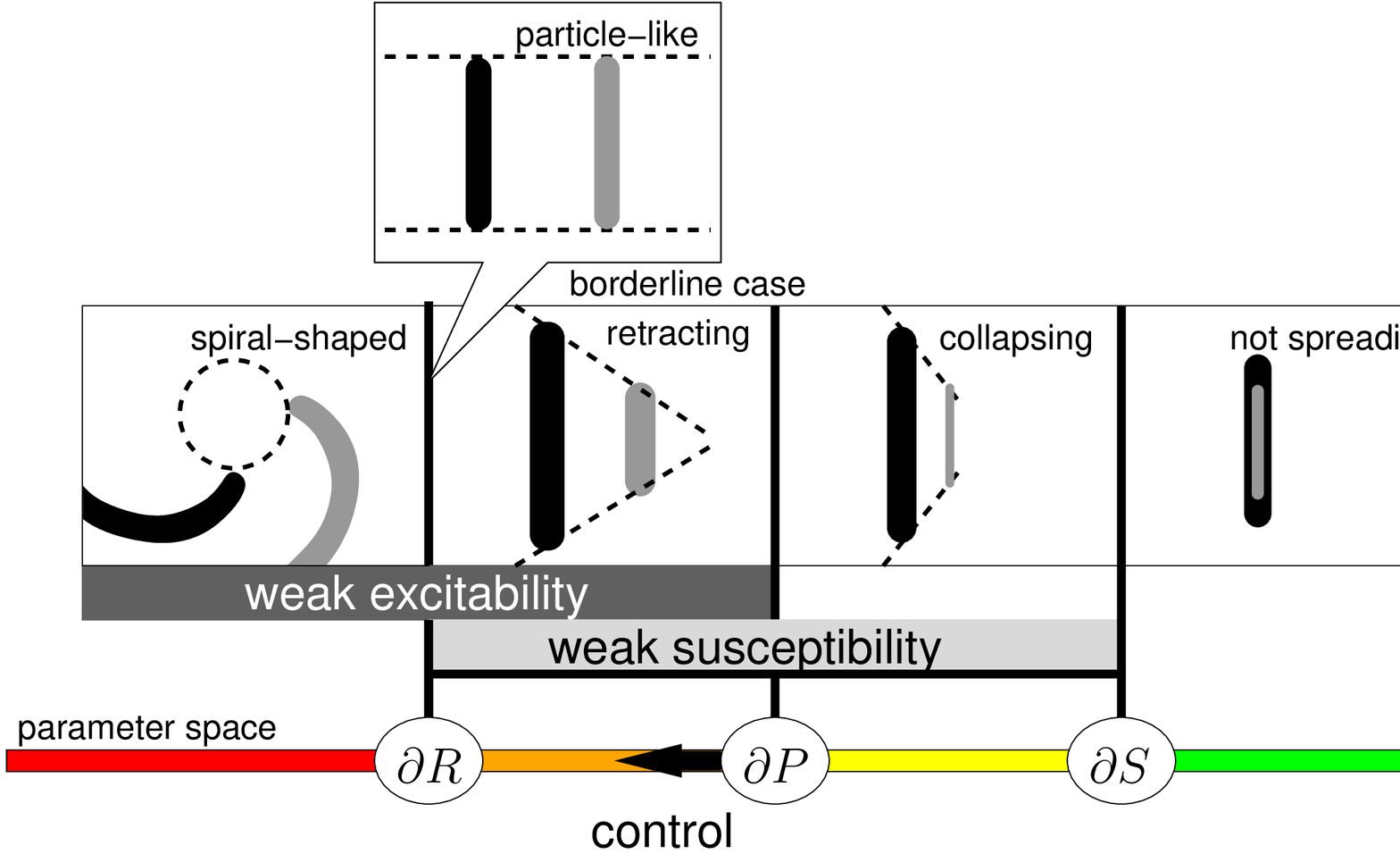}
  \caption{(Color online) Scheme of the classification of excitability according to
  spatio-temporal patterns (spiral-wave, particle-like, retracting,
  collapsing, no spread) in 2D: wave front at two instances in time ($t_1$:
  black, $t_2$: gray) and trajectory of open wave ends (dashed).  This scheme
  illustrates  the effect of control on the propagation boundary ($\partial P$)
  by additional nonlocal and time-delayed transmission capabilities. $\partial
  P$ is controlled by shifting the bifurcation in parameter space (horizontal
  (multi-colored) line).  This affects the regime of weak susceptibility
  centered around $\partial P$ and bounded by the rotor boundary $\partial R$
  and the spreading boundary $\partial S$. } {\label{fig:phases}} \end{figure}

The important criterion for spatial excitability is the existence of a
traveling pulse solution.  In contrast to an excitable local element and
its corresponding phase space configuration (inset in
Fig.~\ref{fig:class}), excitability in active media is based on bistability.  A
superthreshold stimulation takes the system from the homogeneous steady state
into the basin of attraction of the pulse solution.  The classification of
spatial excitability in active media is based on this configuration in phase
space.  For example, the primary rough classification into excitable and
non-excitable media is based on the existence and non-existence, respectively,
of the pulse as the most basic spatio-temporal pattern that sustainedly
propagates in space.  This \textit{propagation boundary} is  called $\partial
P$ \cite{WIN91}.  The focus in this study is on the onset of excitability in
active media at $\partial P$.

The propagation boundary $\partial P$ is determined in the parameter space of a
co-moving frame. In this frame, $\partial P$ is caused by a saddle-node
bifurcation above which traveling pulse solutions exist. As active media are
usually based on a reaction-diffusion mechanism, this bifurcation is computed
in a parabolic partial differential equation (PDE).  The details of this
computation can be found in textbooks \cite{KUZ95} and have been further
related to the phenomenon of SD in the cortex  in Ref.~\cite{DAH08}.
Therefore, we just sketch the basic idea.  The propagation boundary in a
parabolic PDE is obtained by searching pulse profiles as stationary solutions
in a co-moving frame. The profile equation in the co-moving frame must tend to
the fixed point value of the original system for $\xi \rightarrow \pm \infty$,
with $\xi$ being the spatial coordinate in the co-moving frame.  A traveling
pulse is thus equivalent to the existence of a homoclinic orbit satisfying a
system of ordinary differential equations, namely the profile equation
system, with appropriate boundary conditions.

In Sec.~\ref{sec:spcg} we describe how the locus of the propagation boundary
$\partial P$ in the parameter space is controlled by nonlocal spatial coupling
and time-delayed feedback by a shift to new parameter values without adding a
distinctly new character to the spatio-temporal patterns, as schematically
illustrated in Fig.~\ref{fig:phases}.  Since the propagation boundary is
essentially a feature of an active medium with one spatial dimension, we can
limit our main investigation to one-dimensional systems.  Yet, to get a picture
of the patterns that arise in the neighborhood of $\partial P$, we will end
Sec.~\ref{sec:lese} by a brief review of patterns in higher spatial dimensions,
and provide definitions of   {\em weak excitability} and {\em weak
susceptibility}.

\subsection{Weak excitability and weak susceptibility}
\label{sec:wews}

The variety of qualitatively different spatio-temporal patterns in higher
excitable regimes, i.\,e., beyond $\partial P$ towards the oscillatory regime
(to the left of $\partial P$ in Fig.~\ref{fig:class}), provides the foundation
for a classification of spatial excitability in active media. This is in so far
analogous to the classification into type I and II of excitable local elements
(Sec.~\ref{sec:le}) as both classifications are built on the patterns that
emerge. However, the classification of local excitability defines different
types of mechanisms, whereas the  classification of spatial excitability
characterizes rather the degree of excitability.

Excitability is described by ordinary differential equations in local elements
and,  by partial differential equations in spatial media. Roughly speaking,
the additional independent spatial variable allows for more complex patterns,
due to the fact that the spatial dimension renders the phase space
infinite-dimensional. In fact, the patterns can become even more complex if
spatial dimensions are increased from one to two (for example spiral waves in
retinal spreading depression \cite{DAH97}) or three dimensions (Winfree
turbulence of scroll waves in cardiac fibrillation \cite{WIN94a}).

Close to the propagation boundary, the complexity of emerging patterns is
largely independent of the number of spatial dimensions. This fact is also
paraphrased as {\it weak excitability} \cite{MIK91}.  This term is used for
active media to indicate that either no reentrant patterns occur (described by
the rotor boundary $\partial R$ \cite{WIN91}, see Figs.~\ref{fig:class} and
\ref{fig:phases}) or that the rotation period is large enough, so that the
front of the pulse does not interact with its refractory back.  The onset of
interactions between front and refractory back in a reentrant wave pattern is
described by the meandering boundary $\partial M$, see Fig.~\ref{fig:class}. To
the left of this boundary, the core of a freely rotating spiral wave performs a
meandering pattern \cite{NAG93,BRA93}, whereas to the right of $\partial M$ the
spiral core follows a rigidly circular rotation \cite{SCH06c}(Fig.~\ref{fig:phases}).
Changing excitability parameters further, the spiral core can start to perform
more complex manoeuvres to avoid the  refractory zone (complex boundary
$\partial C$), see Fig.~\ref{fig:class}. 
 
Patterns of spreading depression in chicken retina are observed in vitro in the
complex regime to the left of $\partial C$ \cite{DAH97}, but in human brain
tissue they occur close to $\partial R$.  It was predicted that the window of
cortical excitability lies between $\partial R$ and $\partial P$ \cite{DAH04b}
and this seems to be confirmed by a functional magnetic resonance imaging study
in migraine, mapping spatio-temporal patterns of symptom reports onto the
folded cortical surface \cite{DAH08d}.  These patterns are similar to
particle-like or retracting wave segments that occur between $\partial R$ and
$\partial P$ \cite{SAK02, ZYK05, MIK06}.  Transient patterns can also be
observed at even lower excitability, limited by the spreading boundary
$\partial S$ \cite{DAH07a} (Fig.~\ref{fig:phases}).  Since the regime of
retracting wave segments is not identical with weak excitability (absence of
front-back interactions) it was called weak susceptibility based on a
susceptibility scale that can also be operationally defined in experimental
systems \cite{DAH08d}.  The control of excitability in media being weakly
susceptible to pattern formation can be investigated in a model with one
spatial dimension, because $\partial P$ is essentially a property of a 1D
parabolic PDE.

\section{Reaction-diffusion with augmented transmission capability}
\label{sec:rd}

The generic framework to model spatial excitability is  a reaction-diffusion
system of activator-inhibitor type. In fact, already a single species model
has the capacity to propagate fronts \cite{SCH72}. Such a model was originally
suggested for SD by Hodgkin and Grafstein \cite{GRA63} (Sec.~\ref{sec:HG}). We
extend this generic model by choosing appropriate inhibitor dynamics
(Sec.~\ref{sec:extendInhibitor}) to obtain pulses with the onset saddle-node
bifurcation $\partial P$  and introduce augmented transmission capabilities
(Sec.~\ref{sec:augmentedTransmissionCapability}).

\subsection{Hodgkin-Grafstein model}
\label{sec:HG}

A single species model of bistability has the capacity to propagate one state
into the other if these bistable dynamics of the species, called $u$, are
coupled by diffusion 
\begin{eqnarray} 
 \label{eq:singleSpecies} 
 \frac{\partial u}{\partial t}&=&f(u)\; - v \;+\;\frac{\partial^2 u}{\partial x^{2}},\\
 \label{eq:cubic} 
 f(u)&=&u - \frac{u^3}{3}.
\end{eqnarray} 
Here $v$ is a bifurcation parameter. In these non-dimensional equations the
diffusion coefficient, which merely scales space, is set to unity. 

Eqs.~(\ref{eq:singleSpecies})-(\ref{eq:cubic}) were proposed by Hodgkin as a
model for spreading depression based on the bistable $K^+$ dynamics suggested
by Grafstein \cite{GRA63}.  The front profile of the species $u$ and its speed
can be calculated analytically  for the cubic polynomial form of $f(u)$, i.,e., the
generic form of bistable dynamics (see for example the textbook
references~\cite{WIL99,SCH00}).  In the context of control one should think of
the two stable states as one being a physiological (healthy) state and the
other one being a pathological (depolarized) state, and the latter invading the
former, which shall be prevented by control.

As is shown in Sec.~\ref{sec:spcg}, control of the onset of propagation depends
essentially on the interaction of the healthy state with the pulse front via
the augmented transmission capability.  However, in the system defined by
Eqs.~(\ref{eq:singleSpecies})-(\ref{eq:cubic}) there does not exist a boundary like
$\partial P$ (Sec.~\ref{sec:se}). In other words, there is no abrupt onset of
the capacity to propagate fronts with finite speed.  Instead, the heteroclinic
orbit, which corresponds to the front profile in a co-moving frame, exists  for
the whole bistable regime,  in particular also for  arbitrarily slow velocities
of the co-moving frame including zero (standing front) \cite{WIL99}.  To obtain
propagation onset behavior, we need to add inhibitor dynamics, i.\,e., a
second recovery species for the Hodgkin-Grafstein model.

\subsection{Extension by inhibitor dynamics} \label{sec:extendInhibitor}

With a proper choice of a second species whose dynamics is coupled to $u$, the
local system can change from bistable to excitable dynamics, which is needed to observe
and compute $\partial P$.  The natural choice is to use the bifurcation
parameter $v$ of the Hodgkin-Grafstein scheme as the additional species, namely
as an \textit{inhibitor}:  
\begin{eqnarray} \label{eq:inhibitor_0}
  \frac{\partial v}{\partial t}&=&\varepsilon g(u,v).  
\end{eqnarray} 
The rate equation $g(u,v)$ of the inhibitor determines the dynamics in the
refractory phase of the pulse, where it recovers the initial healthy state that was
recruited into the pathological  state of species $u$. In this two-species
model, $u$ is called the \textit{activator}.  Inhibitor dynamics usually changes
on a slower time scale $\varepsilon \ll 1$.  

There are several models in the neuroscience literature obeying this structure
in Eqs.~(\ref{eq:singleSpecies})-(\ref{eq:inhibitor_0}), both models with local
excitability  of type I \cite{HIN82,WIL99} and of type II
\cite{BON53,FIT61,NAG62}.  These models are, however, models of action
potentials describing the propagation of normal electrophysiological activity
along  single nerve fibers. Therefore these models do not relate directly to
SD, because SD is caused by large-scale neuronal activity \cite{KAG00,FLO08}.
Therefore the analogy to models as in Refs.~\cite{HIN82,WIL99,BON53,FIT61,NAG62}
is mainly formal in the mathematical structure  but not in its biological
interpretation.  For example, in the Hodgkin-Grafstein
Eqs.~(\ref{eq:singleSpecies})-(\ref{eq:cubic}) the activator is $K^+$ while in
action potential models it is the membrane potential.

Models exhibiting type I excitability are also capable to show type II behavior
for a certain parameter regime but the reverse is not necessarily true.  This
suggests that type II behavior is more generic, because it can be observed in
both types of models and only requires a rate function $g(u,v)$ that is linear
in both arguments.   We introduce inhibitor dynamics
as
\begin{eqnarray}
  \label{eq:inhibitor}
g(u,v) &=& u + \beta - \gamma v.
\end{eqnarray}
With this choice, the local excitable dynamics is of type II.  The
parameter $\beta$ and $\gamma$ determine the precise configuration of
trajectories in phase space that lead to excitability and are thus, in addition
to $\varepsilon$, the only two parameters that control excitability of type II
in a normal form model.

Eqs.~(\ref{eq:singleSpecies})-(\ref{eq:inhibitor}) correspond to the well
studied FitzHugh-Nagumo (FHN) systems, which is, in fact, used as a paradigm
for excitable systems \cite{LIN04}. Pattern formation in a net of
FitzHugh-Nagumo elements  under the influence of time-delayed feedback was
investigated with the aim to increase the coherence of noise-induced wave
patterns \cite{GAS07b,GAS08}.

\subsection{Augmented transmission capability}
\label{sec:augmentedTransmissionCapability} 

\begin{figure}[tb]
\includegraphics[width=0.4\textwidth]{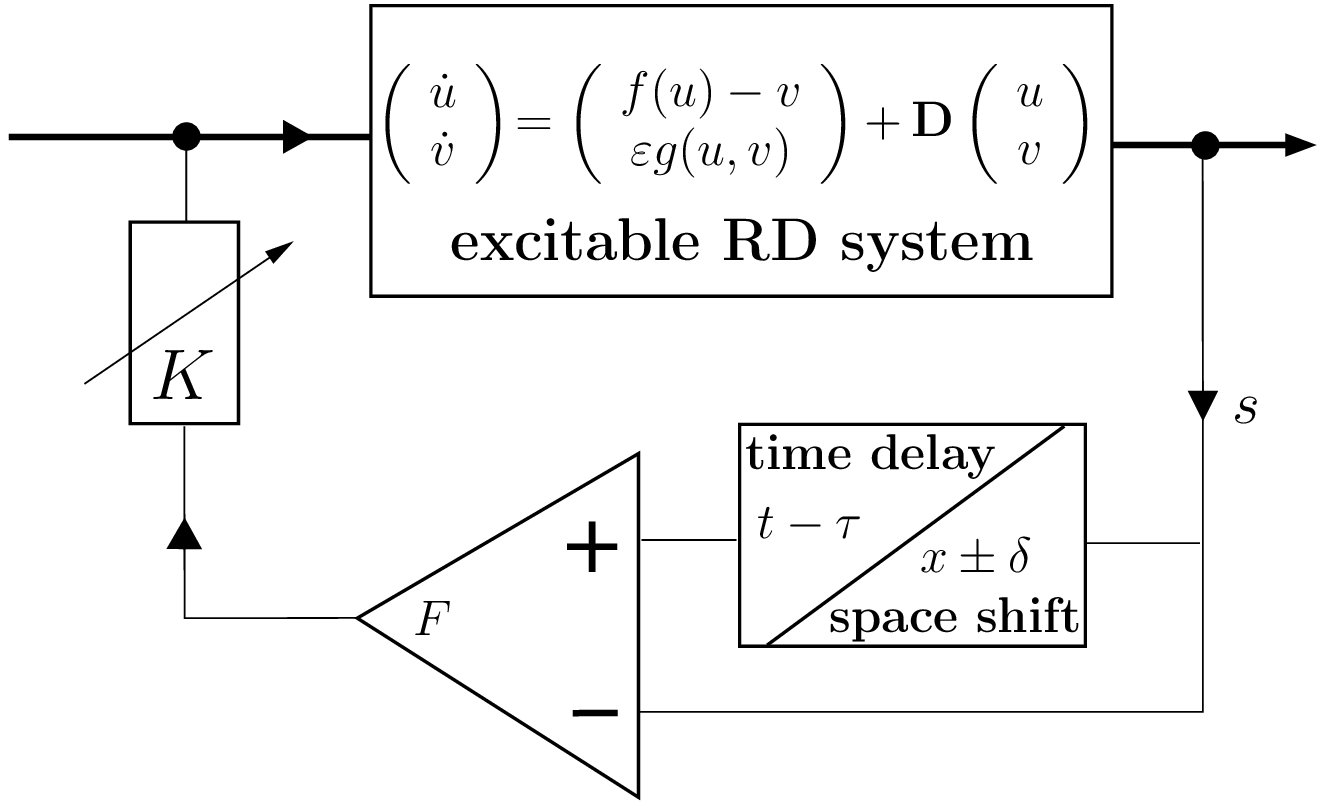}
\caption{
Diagrammatic view of control by augmented transmission (CAT).}
{\label{fig:controlScheme}}
\end{figure}

We extend the FHN model by introducing augmented transmission capability into
Eqs.~(\ref{eq:singleSpecies})-(\ref{eq:inhibitor}) as a feedback loop \cite{Felix}
\begin{eqnarray}
\label{eq:main}
\left(\!
\begin{array}{*{1}{c}}
\partial_t u\\
\varepsilon^{-1}\partial_t v
\end{array}
\!\right)=
\left(\!
\begin{array}{*{1}{c}}
 f(u)-v\\
g(u,v)
\end{array}
\!\right) + 
{\bf D}\left(\!
\begin{array}{*{1}{c}}
u\\
v
\end{array}
\!\right)+ 
 \,\bf{H}\left(\!
\begin{array}{*{1}{c}}
u\\
v
\end{array}
\!\right),
\end{eqnarray}
where $\bf {D}$ is the local diffusion operator, and $\bf{H}$ represents the
augmented transmission capability.
In Eqs.~(\ref{eq:singleSpecies})-(\ref{eq:inhibitor})
we have considered diffusion in the activator species $u$ only:
\begin{eqnarray} 
{\bf D}=\left(\begin{array}{*{2}{c}} \nabla^2 & 0\\ 0 & 0 \end{array} \right).
\end{eqnarray}
The augmented transmission capability
\begin{eqnarray}
  \label{eq:H} {\mathbf{H}}=K\, {\mathbf{F}} 
 \end{eqnarray}
is described by the control strength K and the control matrix
 
\begin{eqnarray}
  \label{eq:F} {\mathbf{F}}= \left(
  \begin{array}{*{2}{c}}  F_{uu} & F_{uv} \\ F_{vu}  & F_{vv}  \end{array}
    \right), 
 \end{eqnarray}
whose elements $F_{ij}$ are operators which represent three individual steps of
the control by augmented transmission (CAT), namely (i) selecting a species $j$
whose transmission capability is augmented, (ii) creating the control force from this
species, and (iii) feeding this control force back into the dynamical variable
$i$, see Fig.~\ref{fig:controlScheme}. 

Formally, this can be represented by 
splitting \mbox{$ F_{ij}=A_{ij}F$} into the components of a coupling matrix
${\mathbf A}$, which represents the coupling scheme, e.g.,
\begin{eqnarray}
\label{eq:A}
{\mathbf A}^{uu}=\left(\begin{array}{*{2}{c}} 1 & 0\\ 0 & 0 \end{array} \right),\quad
{\mathbf A}^{uv}=\left(\begin{array}{*{2}{c}} 0 & 1\\ 0 & 0 \end{array} \right),\nonumber \\
{\mathbf A}^{vu}=\left(\begin{array}{*{2}{c}} 0 & 0\\ 1 & 0 \end{array} \right),\quad
{\mathbf A}^{vv}=\left(\begin{array}{*{2}{c}} 0 & 0\\ 0 & 1 \end{array} \right),
\end{eqnarray}
and an operator $F$ that creates the type of control signal.

The control signal $F$ is generated from the input variable $s=u$
or $s=v$ by time-delayed or nonlocal feedback.
For example, in time-delayed feedback as introduced by Pyragas
\cite{PYR92},  the operator $F$ creates the difference between the
time-delayed signal $s(t-\tau)$ and its current counterpart $s(t)$. We study
this type and two nonlocal types of coupling. The control force  $F$ of these
different types, as a function of the signal $s$, is described in detail in
Sec.~\ref{sec:nonlocal_control} and \ref{sec:TDAS}. 
It should be noted that in Eq.~(\ref{eq:main}), ${\bf H}$ represents cortical
circuits and neurovascular coupling \cite{DAH08}, that is, {\it internal}
control loops.

The four coupling matrices in Eq.~(\ref{eq:A}) lead to four transmission
pathways, also termed {\em coupling schemes}: two self-couplings $uu$ and $vv$,
and two cross-couplings $vu$ and $uv$, corresponding to the indices $ij$ 
in Eq.~(\ref{eq:F}). 
Note that other coupling schemes, for example diagonal
coupling or coupling with a rotation matrix \cite{KYR09,FIE07,DAH07}, can also
be investigated within this framework. 

\section{Suppression of pulse propagation}
\label{sec:spcg}

We perform simulations of the RD system in
\mbox{Eqs.~(\ref{eq:singleSpecies})-(\ref{eq:inhibitor})} in one spatial
dimension with  RD parameter settings ($\varepsilon$=0.1, $\beta$=0.85,
$\gamma$=0.5) for the weakly excitable regime close to the propagation boundary
$\partial P$ (Sec.~\ref{sec:se}). The RD system is extended by three types of
coupling in the framework of Eq.~(\ref{eq:main}), i.e., two types of nonlocal
spatial coupling (isotropic and anisotropic) and one type of time-delayed
feedback.  In the context of spreading depolarizations, nonlocal spatial and
local time-delayed couplings represent neural structural and functional
connectivity and neurovascular feedback in the cortex. This is discussed in
more detail in Ref.~\cite{DAH08}.  For each type of coupling there are four
principal coupling schemes, two self-coupling schemes $(uu,vv)$ and two
cross-coupling schemes $(uv,vu)$, defined by the coupling matrix ${\mathbf A}$
in Eq.~(\ref{eq:A}).

We initialize each simulation with a stable pulse profile solution of the RD
system in Eqs.~(\ref{eq:singleSpecies})-(\ref{eq:inhibitor}) to investigate the
effect of control on  RD excitability.  This is accomplished by testing whether
this specific initial condition of the free system (RD only) is in the basin of
attraction of the homogeneous steady state of Eq.~(\ref{eq:main}), i.\,e., the
controlled system (RD-CAT).  There are two CAT parameters: The control gain $K$
in Eq.~(\ref{eq:H}), and  a  spatial ($\delta$) or temporal
($\tau$) control scale, repectively, which will be introduced in the following
sections.  For a large range of these two CAT parameters, we determine whether
pulse propagation is suppressed or not. The propagation is suppressed if the
excitation dies out, so that the system approaches the homogeneous steady
state.  In this case, CAT is considered successful.  In the reverse case,
any sustained spatio-temporal pattern that evolves from the initial conditions
(free pulse solution), after the CAT is "switched on",  is considered an
unsuccessful control since the activity is not completely suppressed and the
homogeneous steady state is not reached. 

For the following simulations we adopt an active medium with a spatial
extension of $L=160$. As spatial increment in the discretized Laplacian
$\mathbf D$ we take $\delta x=0.2$.  All simulations are run for 2000 time
units and use an Euler forward algorithm with discretization $\delta
t=0.00125$. The spatial and temporal width of the free-running activator pulse,
$\Delta x$ and $\Delta t$, respectively, serve as reference space and time
scales \cite{DAH08}.

\subsection{Pulse suppression by nonlocal spatial coupling}
\label{sec:nonlocal_control}

In this section we present results on two types of spatial coupling.

\begin{figure}[b!]
\includegraphics[width=0.48\textwidth]{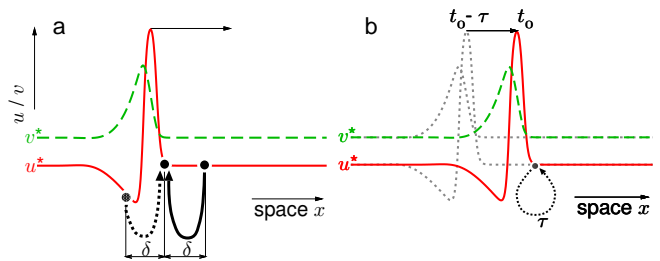}
\caption{(Color online) Schematic diagram illustrating nonlocal spatial coupling and
time-delayed feedback. (a) Nonlocal spatial coupling: isotropic coupling
(dashed and solid arrow), and anisotropic  backward coupling (solid arrow). (b)
Time-delayed feedback: backward coupling of traveling pulses. 
Pulse profiles are shown in red (solid) and green (dashed)
for the activator and inhibitor, respectively. Dotted: time-delayed
profiles.}
{\label{fig:coupling}}
\end{figure}

\subsubsection{Isotropic coupling}
\label{sec:spsmcg}
Nonlocal isotropic spatial coupling is defined as  
\begin{equation}
  \label{eq:isotropic_coupling}
  F(s)=s(x+\delta,t)+s(x-\delta,t)-2s(x,t)
\end{equation}
where $s=u$ or $s=v$. 

\begin{figure}[!tb]
\includegraphics[width=0.48\textwidth]{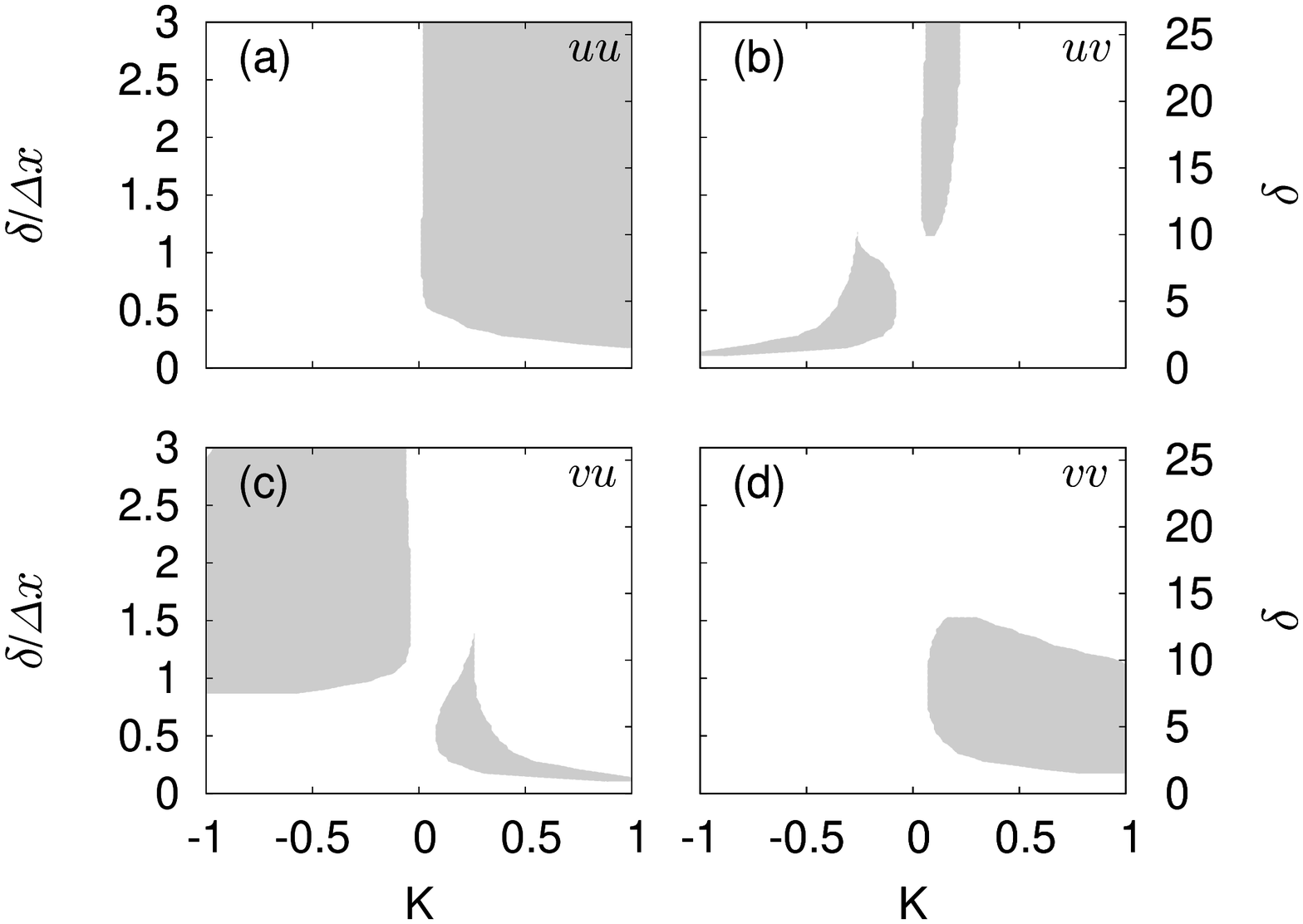}
\caption{
Control planes of isotropic spatial coupling spanned by the two CAT parameters:
control gain factor  $K$ and nonlocal space scale $\delta$ normalized to pulse
width $\Delta x$ (left scale) and in spatial units (right scale).  (a)
Activator self-coupling scheme $uu$, (b) cross-couplings $uv$ (inhibitor signal
fed back to activator rate equation) and (c) $vu$ (reverse), and (d)
inhibitor self-coupling $vv$. Suppression of pulse propagation is marked by
gray control domains. Parameters: $\varepsilon=0.1$, $\beta=0.85$, $\gamma=0.5$.
$\Delta x=8.65$. }
{\label{fig:isotropic_coupling}}
\end{figure}

Regimes in which reaction-diffusion pulses are suppressed by additional
nonlocal connections are calculated for the parameters $\delta$ and $K$ and
represented by gray areas in Fig.~\ref{fig:isotropic_coupling}.   For the two
self-coupling schemes $uu$ and $vv$, shown in Fig.~\ref{fig:isotropic_coupling}
(a) and (d), respectively, the sign of the gain parameter $K$ determines the
effect of the nonlocal connection.  Pulse propagation can only be suppressed
for $K>0$.  For the two cross-coupling schemes $uv$ and $vu$, shown in
Fig.~\ref{fig:isotropic_coupling}  (b) and (c), respectively, the sign of the
gain parameter $K$ changes at  $\delta\approx \Delta x$ for  pulse suppression,
and these two cross-coupling schemes show similar control domains with respect
to reflection $K \rightarrow -K$. 

The sign of $K$ in the control domains of the self-coupling schemes
(Fig.~\ref{fig:isotropic_coupling} (a) and (d)) can be qualitatively understood
by considering the effect of nonlocal connections upon the homogeneous steady
state in the limit $\delta \rightarrow 0$.  For $K>0$, this limit corresponds
to diffusively coupled elements.   In general, the homogeneous steady state is
stabilized by diffusion against small inhomogeneous perturbations.  In the same
way, a local perturbation is leveled by  a nonlocal connection in the form of
Eq.  (\ref{eq:isotropic_coupling}) for self-coupling.  In the diffusion limit,
however, $K$ would increase the diffusion coefficient, which causes the pulses
to become broader.  Yet, a qualitative change in the dynamics cannot occur by
changing the diffusion coefficient. Therefore the effect of suppressing pulses
depends on the nonlocal character of the connection, which must extend at
least over a distance of  about 20\% of the pulse width $\Delta x$.

The control domains of  the cross-coupling schemes
(Fig.~\ref{fig:isotropic_coupling} (b)-(c)) are qualitatively different from
self-coupling.  $K$ changes its sign at  $\delta\approx \Delta x$ within one
scheme and accordingly the situation of  pulse suppression is more complex.
There is a  long-range regime ($\delta > \Delta x$) and a short-range regime
($\delta < \Delta x$).  The sign of $K$  for pulse suppression in the
long-range regime depends mainly on the backward connection (solid arrow in
Fig.~\ref{fig:coupling}(a)), as will be shown in the next section. In the
short-range regime, the effect of the forward connection (dotted arrow in
Fig.~\ref{fig:coupling}(a)) seems to dominate and reverse the effect observed
in the long-range regime.

\subsubsection{Anisotropic backward coupling}
\label{sec:spascg}

Nonlocal connections can also be introduced in only one direction 
\begin{equation}
  \label{eq:anisotropic_backward_coupling}
  F(s)=s(x\pm\delta,t)-s(x,t).
\end{equation}
This directed connectivity would correspond to anisotropic nonlocal
coupling in a two dimensional excitable medium. One reason to investigate this
type of connectivity is to obtain a better understanding of the results in the
previous section by separating effects of forward  and backward connections
(Fig.~\ref{fig:coupling} (a)). Moreover, the functional and structural
connectivity of the cortex is realistically modeled as an anisotropic (and also
inhomogeneous) medium due to the patchy nature of nonlocal horizontal cortical
connections. While anisotropies in the cortical connections are usually
considered  to merely cause variations in wave speed in different directions,
inhomogeneities are known to cause wave propagation failure \cite{BRE01}. In
contrast, our focus is on the change in excitability by anisotropic nonlocal
coupling that leads to suppression of wave propagation.

We limit your investigation to the backward connection.  This corresponds to
the plus (minus) sign in Eq.~(\ref{eq:anisotropic_backward_coupling}) for
pulses propagating in the positive (negative) $x$ direction (solid  arrow in
Fig.~\ref{fig:coupling}(a)). We choose the backward connection because  this
type of coupling shares many properties with  local time-delayed feedback
coupling, as we will show by comparing the results from this  section  with
those of the next section (Sec.~\ref{sec:TDAS}).

\begin{figure}[b]
  \centering
   \includegraphics[width=0.48\textwidth]{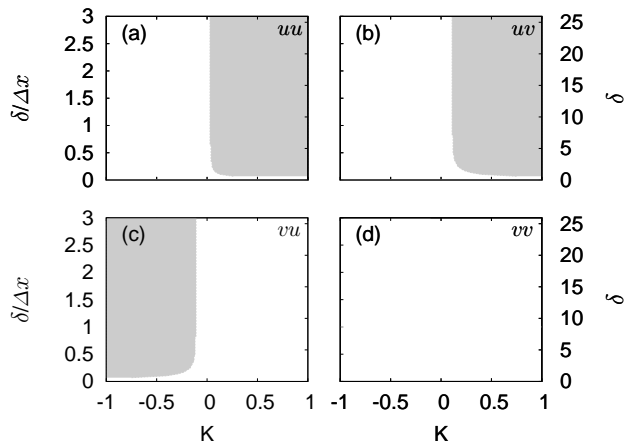}
   \caption{
   Control planes of {\it anisotropic} backward spatial coupling.
   Parameters and notation as in Fig.~\ref{fig:isotropic_coupling}. 
   \label{fig:spacedelay2}}
\end{figure}

The control domain of the $uu$ self-coupling scheme (Fig.~\ref{fig:spacedelay2}
(a)) is very similar to the isotropic one shown in
Fig.~\ref{fig:isotropic_coupling} (a). This indicates that in  the $uu$ scheme
of isotropic coupling the  backward connection accounts for  the main
contribution to suppression of wave  propagation. However, we want to note that
both control types (isotropic and anisotropic) with the same coupling scheme
$uu$ can differ in their efficiency within some parts of the gray control
domain (not shown).  The efficiency refers to the length a pulse travels  after
the connectivity is "switched on".  This transient effect was investigated in
detail in \cite{DAH07a,DAH08}, whereas in this study we concentrate on an
understanding of the size and shape of control domains for the different
schemes and types of coupling
(Figs.~\ref{fig:isotropic_coupling}-\ref{fig:controldomains}).

Pulse propagation is not suppressed by the $vv$ self-coupling scheme for
anisotropic backward coupling for any control parameter within the investigated
range (Fig.~\ref{fig:spacedelay2}(d)).  This, however, does not mean that the
pulse propagates essentially unchanged.  Any sustained spatio-temporal pattern
other than the homogeneous steady state that evolves from the free initial
pulse solution---after the connectivity is "switched on"---is considered as
unsuccessful pulse suppression.  In fact, for the $vv$ scheme of anisotropic
coupling in the expected control domain of corresponding
Fig.~\ref{fig:isotropic_coupling} (d), the homogeneous steady state becomes
unstable and stationary spatial patterns emerge after the initial pulse is
suppressed.

The control domains of the cross-coupling schemes (Fig.~\ref{fig:spacedelay2}
(b)-(c)) are similar to Fig.~\ref{fig:isotropic_coupling} (b)-(c) in the
long-range regime, except for a failure of suppression in the isotropic
$uv$ scheme for $K>0.2$.  The long-range regime for anisotropic
backward cross-coupling extends over the nearly complete  spatial scale of
$\delta$ (except for very small $\delta$).  Thus, $K$ does not change its
sign. Also, there is a perfect reflection symmetric pattern in the
control planes with respect to the axis $K=0$ for the two cross-coupling
schemes. 

The reason for the reflection symmetry in the pattern of control domains for
cross-coupling schemes and, furthermore, the explanation for the observed signs
of the gain parameter $K$ for the control domains of all coupling schemes for anisotropic backward
coupling are given in Sec.~\ref{sec:effectiveParameters} together with
the explanation of some of the results we have obtained for local time-delayed feedback
control (Pyragas feedback), which will be described  in the next section.

\subsection{Pulse suppression by local time-delayed feedback}
\label{sec:TDAS}

\begin{figure}[t]
\includegraphics[width=0.48\textwidth]{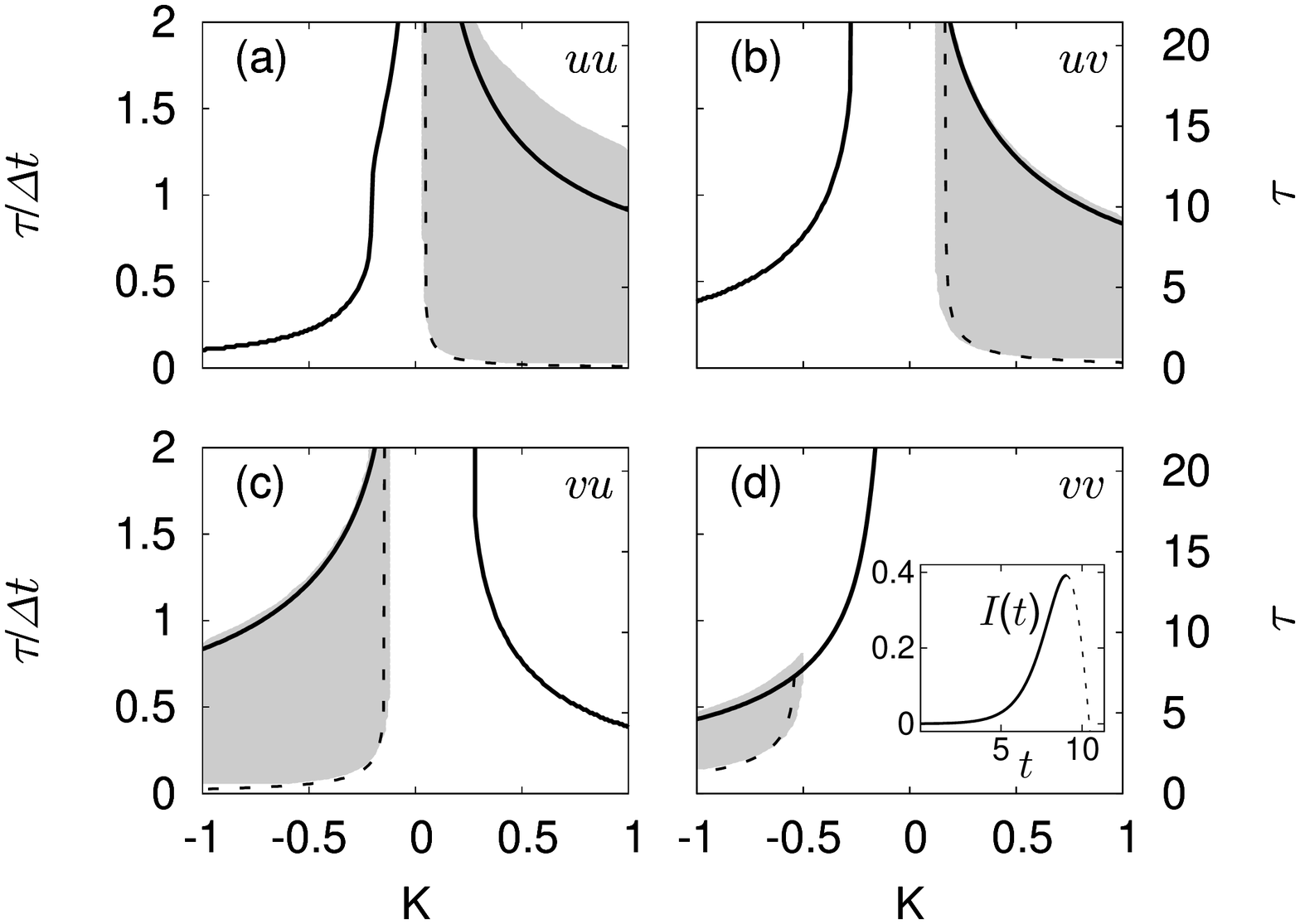}
\caption{
Control planes of time-delayed feedback coupling spanned by the two CAT
parameters: control gain factor  $K$ and time delay $\tau$ normalized to pulse
width $\Delta t$ (left scale) and in temporal units (right scale).  (a)
Activator self-coupling scheme $uu$, (b) cross-couplings $uv$ (inhibitor signal
fed back to activator rate equation) and (c) $vu$ (reverse), and (d)
inhibitor self-coupling $vv$. Suppression of pulse propagation is marked by
gray control domains. The black lines denote saddle-node bifurcations.
The dashed lines mark the values of control parameters
for which the $u$-amplitude of a single local FHN system,
stimulated in $u$ by $I(t)$ (shown as inset in (d)), is decreased by $10 \%$
compared to the $u$-amplitude of the uncontrolled system.
Parameters as in Fig. \ref{fig:isotropic_coupling}, $\Delta t=10.73$.}
{\label{fig:controldomains}}
\end{figure}

In case of local time-delayed feedback the control force $F$ is
given by 
\begin{equation}
  \label{eq:TDAS}
  F(s)=s(x,t-\tau)-s(x,t),
\end{equation}
where $\tau$ is the delay time.
Note the formal similarity to  Eq.~(\ref{eq:anisotropic_backward_coupling}).
This control method was first introduced by Pyragas \cite{PYR92} for chaos
control.  In Fig.~\ref{fig:coupling} (b), this control method is illustrated for
the $uu$ coupling scheme: at each spatial location $x$ the activator
$u$ at time $t$ receives the signal from the same location but at the
past time $t-\tau$. That means particularly for the dynamics of the front 
that the deviation from the homogeneous fixed point is fed back.

\begin{figure}[!tb] \centering 
  \includegraphics[width=0.4\textwidth]{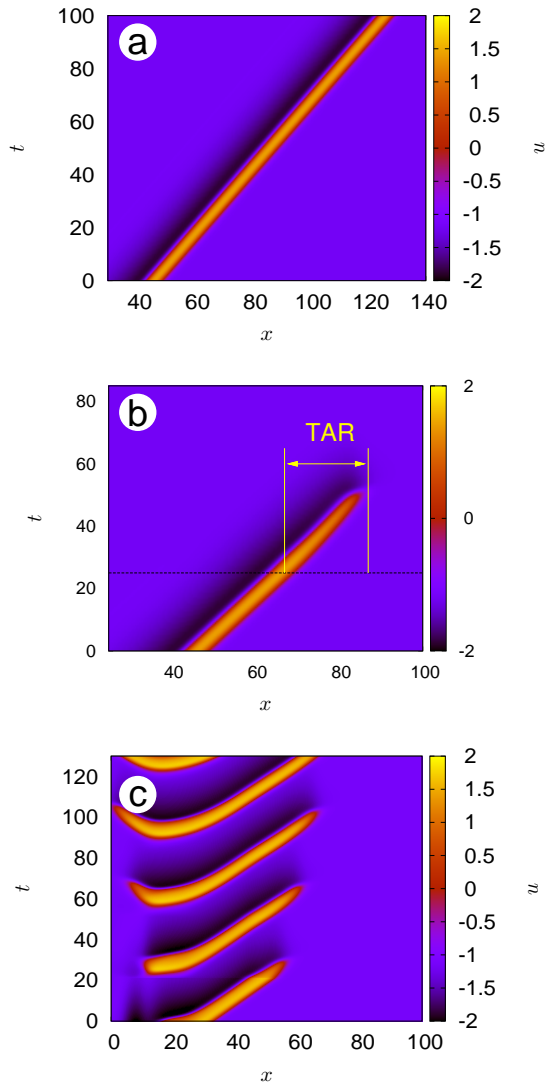}
     \caption{
     Pattern formation for $\varepsilon=0.1$, $\beta=0.85$ and $\gamma=0.5$: (a):
     pulse propagation without control; (b): suppressed pulse via nonlocal
     coupling (\mbox{$K$ = $0.2$}, \mbox{$\delta$ =
     $0.58\Delta x$}).  The horizontal line at $t=25$ marks the onset of control. The
     distance that the transient excitation spreads after control is activated
     before the medium relaxes to the homogeneous fixed defines the transient
     activation radius (TAR) or {\em tissue at risk}.  (c): delay-induced
     oscillatory tracking pattern ($K=0.5$, $\tau=2.8\Delta t$, onset of control at $t=21$)}
{\label{fig:pattern}}
\end{figure}

The domains of successful control, i.e., pulse suppression, in the $(K,\tau)$
plane are shown as gray areas in Fig.~\ref{fig:controldomains}.  For successful
control, the pulse dies out and the system returns to the homogeneous
steady state, as shown exemplarily in the space-time plot of Fig.~\ref{fig:pattern}
(b). 

For time-delayed feedback a domain of successful control exists for each
coupling scheme. As for the spatial coupling schemes in
Fig.~\ref{fig:isotropic_coupling} and \ref{fig:spacedelay2}, the two
cross-coupling schemes $uv$ and $vu$ show a reflection symmetry with respect to
the axis $K=0$.  This symmetry is explained by the effective 
parameters introduced in Sec.~\ref{sec:effectiveParameters}.  Coupling schemes
that feed back the signal to the activator can suppress pulses for
positive $K$. For coupling schemes that feed back the signal to the inhibitor
successful control is  possible for negative values of  $K$. This is similar to
the results of anisotropic backward coupling (cf.\ Fig.~\ref{fig:spacedelay2}).
This similarity is due to the similarity of the signals the pulse receives through
the feedback.  In both cases the wave front gets the feedback from the homogeneous
steady state which is ahead of the wave, in a temporal or spatial sense,
respectively.

The main differences between the control domains of time-delayed feedback and 
spatial anisotropic coupling are that in case of time-delayed feedback the control
domains are limited towards large values of $\tau$ (upper borders of gray domains
in Fig.~\ref{fig:controldomains} (a-d)), and, moreover, that there is a control
domain also for $vv$, the inhibitor self-coupling scheme. The limitation of the
control domains towards large values of $\tau$ is caused by the formation of
tracking patterns \cite{MAN06} (Fig.~\ref{fig:pattern}), which is further
explained in Sec.~\ref{sec:delayInduOsci}.  These patterns emerge by  delay-induced
oscillations. The local dynamics becomes bistable due to a saddle-node
bifurcation (black solid lines, Fig.~\ref{fig:controldomains}).  This is demonstrated
in the next section. 

\subsubsection{Effect of time-delayed feedback in a local excitable system}
\label{sec:single_system}

In this section we investigate the effect of time-delayed feedback on a
single local excitable element. There are two main effects. First, depending on
the sign of the control parameter $K$, the activator amplitude is reduced, which
selects the location of the control domain in the control plane with respect to
the axis $K=0$ (Fig.~\ref{fig:controldomains}). Furthermore, we show that a
minimum amplitude reduction of about 10\% is needed to suppress pulse
propagation when these elements form an active medium (lower bounds of the
control domains).  Second, for too large a value of $\tau$, the local dynamics
becomes bistable (upper bounds of the control domains).

\begin{figure}[bp] 
  \includegraphics[width=0.4\textwidth]{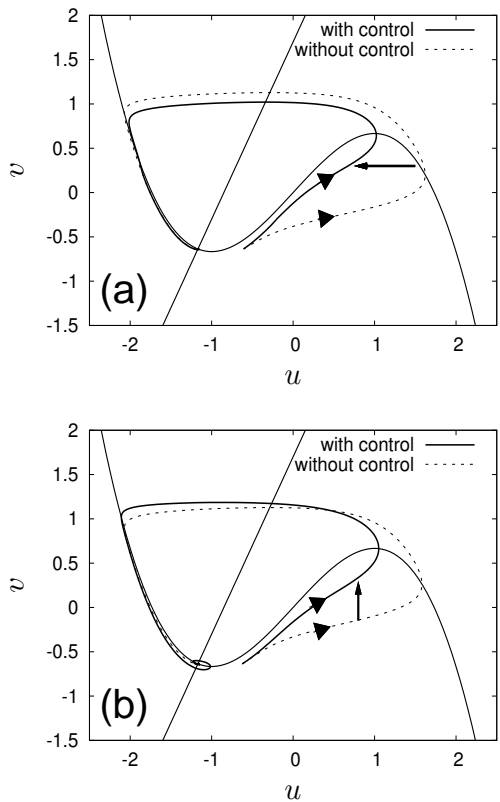}
 \caption{
Phase portraits of a local FHN system with (solid trajectory) and without
(dashed trajectory) time-delayed feedback control.  (a) $uu$ coupling:
K=0.03 (b) $vv$ coupling:  K=-0.7. Parameters: $\varepsilon=0.1$, $\beta=0.85$,
$\gamma=0.5$, $\tau=5$, no diffusion. The solid arrow denotes the action of
the control force.}
  \label{fig:phasenportrait.td}
\end{figure}

To obtain qualitative insight into how time-delayed feedback operates, we 
perform numerical simulations in order to compare trajectories starting from 
the same initial conditions with and without time-delayed feedback. 
Figure~\ref{fig:phasenportrait.td} shows exemplary superthreshold phase space
excursions with and without ($K=0$) time-delayed feedback (solid and dashed 
trajectories, respectively). 
The system is initialized in the interval $\left[ t_0-\tau;t_0 \right[$
(history function) with the fixed point value ($u=u^*$ and $v=v^*$) and for
$t_0$ with a superthreshold value ($u=u^*+0.5$ and $v=v^*$).  

For  coupling schemes that feed back into the activator rate equation,
i.\,e., $uu$ and $uv$, the control force acts parallel to the $u$ axis. In
Fig.~\ref{fig:phasenportrait.td} (a) the trajectories with and without control
for the $uu$ coupling scheme are plotted.  The direction of the control force is
denoted by the horizontal arrow.  In order to reduce the amplitude in $u$, the
control force has to be directed towards the fixed point, i.e. it has to
be negative for $u>u^*$ and $v>v^*$.  This is the case for $t<\tau$ if $K>0$,
because the history function is initialized as $(u^*,v^*)$.  

For the $vu$ and $vv$ coupling scheme the control force acts parallel to the
$v$-axis.  In Fig.~\ref{fig:phasenportrait.td} (b) the trajectories with and
without control are shown for $vv$ coupling.  The direction of the control
force is denoted by the vertical arrow.  To get lower amplitudes of $u$ the
control force has to be directed towards the opposite direction of the fixed
point. This is the case for $t<\tau$ if $K<0$.

In the following the reduction of amplitude $u$ of a single local FHN system is
investigated in order to obtain the lower boundaries (dashed) of the control
domains in Fig.~\ref{fig:controldomains}.
To excite the system a stimulation current $I(t)$ (inset in
Fig.~\ref{fig:controldomains}(d)) is added to the activator $u$.
Choosing for each coupling scheme
the proper sign of control gain $K$ to reduce the activator amplitude,
time-delayed feedback is activated and 
the maximum amplitude $u_{max}$ is observed.
By performing a bisection method for each
value of $K$, the proper $\tau$ is detected that reduces $u_{max}$ by $10\%$
from the original amplitude after stimulation without feedback ($K=0$).  The dashed lines in
Fig.~\ref{fig:controldomains} mark the positions where the $u$-amplitude is
reduced by $10\%$ for a fixed stimulation. These lines form the lower
boundaries of the control domains.

Also the upper boundaries of the control domains can be understood by
investigating the single system with feedback.  For large $K$ and $\tau$ the
system becomes bistable in the sense that in addition to the stable fixed point a
stable limit cycle occurs. This limit cycle appears in a saddle-node
bifurcation of limit cycles, as was also found in a system of two delay-coupled
FHN-systems \cite{DAH08c,SCH08}.  The limit cycle emerges if (i) $\tau$ is
sufficient large to allow for recovery to the fixed point, and (ii) the feedback and
hence the gain parameter $K$ is strong enough to push the system beyond the
threshold.

Again with the help of a bisection method the saddle-node bifurcation lines for
the single system are determined in the $K$-$\tau$ plane.  In
Fig.~\ref{fig:controldomains} they are plotted as black solid lines. These
lines mark the boundaries where the local dynamics becomes bistable and control
fails. Only in case of $uu$ coupling the boundary of the control domain
deviates appreciably from this bifurcation line.  This is due to the
stabilizing effect of diffusion that damps out local oscillations. Since in the
interspace the single system is bistable and thus the medium is able to perform
homogeneous oscillations, the difference between the upper boundary of the
control domain and the bifurcation line depends on the initial conditions that
are chosen to locally stimulate the medium.

Investigating a single FHN-element with time-delayed feedback provides a
qualitative understanding of the dynamics of the controlled spatial system: The
sign of $K$ and the form of the control domains can be understood. Thus the
control of local excitability provides control of the global spatial
excitability.  This will be  quantified in Sec.~\ref{sec:effectiveParameters}.
In the next section, we present the spatio-temporal patterns that emerge above
the upper boundary of the control domain, i.e., the bistable domain of the
single local element.

\subsubsection{Delay-induced oscillatory pattern formation}
\label{sec:delayInduOsci}

As we have seen in the previous section, the local FHN dynamics with
time-delayed feedback becomes bistable for too large values of $K$ and $\tau$.
Due to this local bistability, in the spatially extended system delay induced
oscillatory pattern formation occurs (Fig.~(\ref{fig:pattern} c)): After a local
stimulation the medium performs local oscillations that spread slowly: Within
each local oscillation the excitation spreads a short way, and thus gradually
new parts in the neighborhood become excited. The area of oscillatory
excitation grows slowly all over the medium. We have observed these patterns
for each coupling scheme beyond the upper boundary of the control domain.

This effect will again play an important role in section~\ref{sec:pbs}, 
that focuses on the propagation boundaries in ($\varepsilon,\beta$) - parameter space.

\subsection{Description through effective parameters}
\label{sec:effectiveParameters}

In this section we further invastigate the two control schemes of spatial
anisotropic backward coupling and local time-delayed feedback to achieve an
analytic approximation. In Sec.~\ref{sec:changeYesWeCan} we use this to
describe the change of excitability and in Sec.~\ref{sec:pbs} we investigate
how this is reflected in the parameter space. 

For our approximation we assume that the front dynamics governs the location of
the propagation boundary $\partial P$. Therefore we focus on the part of
control that acts on the front dynamics. For the coupling types of time-delayed
feedback and nonlocal anisotropic coupling the front of the pulse receives the
feedback of the homogeneous steady state. The main idea is to replace the
time-delayed (in Eq.~(\ref{eq:TDAS})) or the space shifted (in
Eq.~(\ref{eq:anisotropic_backward_coupling})) quantities  by their fixed point
values ($u(x,t-\tau)=u(x+\delta,t)=u^*$ and $v(x,t-\tau)=v(x+\delta,t)=v^*$).
This transforms the controlled system, after some simple algebraic
manipulations, into the form of the uncontrolled system (K=0), but with new
\textit{effective} values of the parameters $\tilde{\varepsilon}(K)$,
$\tilde{\beta}(K)$ and $\tilde{\gamma}(K)$.

For each coupling scheme the obtained effective parameters and the
corresponding transformations are shown in the Table~\ref{tab:table}.  Since
the delayed or shifted quantities are replaced by the fixed point values, the
obtained {\it effective} parameters $\tilde \varepsilon$, $\tilde \beta$ and
$\tilde \gamma$ only depend on $K$ and not on $\tau$ or $\delta$. 

Although the transformations look rather complex there is a common feature for
$vv$ and the two cross-coupling schemes $vu$ and $uv$: The change of the
parameters is such that the fixed point value of the
system does not change, i.e., for these three coupling types the fixed
point after the transformation to the system with effective parameters has the
same homogeneous fixed point as the system without control. Since for the $uu$
coupling scheme the transformation is more complex, the fixed point in the
transformed equations differs from that in the original equations, i.e., 
the intersection of the nullclines changes its relative position
with respect to the cubic nullcline.  Note that this does not mean that the
fixed point of the original system changes by adding control.  Due to the
non-invasivity of all investigated control types the fixed point does not change.
However, the transformation that converts the original system with control into
a new system without control but effective parameters, is in case of $uu$ coupling not
invariant for the fixed point ($u^*,v^*$).

\begin{table}
\caption{Effective parameters and transformations for coupling schemes $uu$, $uv$, $vu$,$vv$.}
\label{tab:table}
{\normalsize 
\begin{tabular}{c|l|l|l|l} &$uu$&$uv$&$vu$&$vv$\\
\hline $\tilde t$ & $\frac{t}{1-K}$   &  $ t$  &  $ t $ & $ t $  \\ 
$\tilde x$ &$\frac{x}{\sqrt{1-K}}$ & $x$& $x$&$x$ \\ 
\tabularnewline[-0.8cm]& & & & \\
\hline
\tabularnewline[-0.9cm]& & & & \\
$\tilde u$ &$\frac{u}{\sqrt{1-K}}$ &$u$ &$u$ &$u$ \\ 
\tabularnewline[-0.8cm]& & & & \\
$\tilde v$ & $\frac{(v-Ku^*)}{\sqrt{1-K}^3}$ & $v-K(v^*-v)$&$v$ & $v$\\
\tabularnewline[-0.8cm]& & & & \\
\hline 
\hline $\tilde \varepsilon$ & $\frac{\varepsilon}{(1-K)^{2}}$& $(1+K) \varepsilon$ & $(1-K)
\varepsilon$& $\varepsilon$\\
\tabularnewline[-0.8cm]& & & & \\
$\tilde \beta$ &
$\frac{(\beta-\gamma Ku^*)}{\sqrt{1-K}}$ &$\frac{\beta-Ku^*}{1+K}$ &
$\frac{\beta+Ku^*}{1-K}$&$\beta+Kv^*$\\
\tabularnewline[-0.8cm]& & & & \\
$\tilde \gamma$ &$\gamma (1-K)$&$\frac{\gamma}{1+K}$ &$\frac{\gamma}{1-K}$ &$\gamma+K $\\
\end{tabular} } \end{table}

The effects of parameter changes in the FHN system are well known
\cite{DAH07a,DAH08}.  As rule of thumb one can keep in mind that the 
larger $\varepsilon$,$\beta$ or $\gamma$, the lower the excitability of the system.

In the following the $K$ dependence is discussed:\\
In case of $vv$ coupling $\varepsilon$ does not change.  For
$K<0$ only $\beta$ becomes larger while $\gamma$ decreases. However, the change
of $\beta$ dominates, i.e., the excitability decreases for
$K<0$.

For $vu$ coupling $\tilde{\varepsilon}$ and $\tilde{\beta}$ increase for
$K<0$, while $\tilde{\gamma}$ decreases. In that case the influence of
changing $\tilde{\varepsilon}$ dominates, i.e., excitability
decreases for $K<0$. 

For $uu$ and $uv$ coupling $\tilde{\varepsilon}$ and $\tilde{\beta}$
increase for $K>0$, while $\tilde{\gamma}$ decreases. The influence of
$\tilde{\varepsilon}$ is dominating, and therefore the excitability decreases
for positive values of $K$. 

For the $vv$ and $vu$ schemes the inhibitor receives a feedback. In these cases
by simply rearranging the inhibitor equation the original form without control,
but with the new effective parameters, can be obtained, and therefore no
transformation in time and space or of $u$ and $v$ variable has to be
performed.  For the two coupling schemes, for which the activator receives a
feedback ($uu$ and $uv$) the equations need to be transformed in order to
retain the original form without additional control force.  In case of $uv$
only the $v$ variable is transformed, whereas in case of $uu$ coupling both
dynamic variables $u$ and $v$ and also time and space are transformed. However,
this does not change the qualitative dynamics of the system, since
transformations in $u$ and $v$ variable and in time and space correspond to
rescaling only. 

For the two cross-coupling schemes the effective parameters are symmetric with
respect to $K\rightarrow-K$. This symmetry of the cross-coupling schemes
was observed in the control domains of all control types
(Figs.~\ref{fig:isotropic_coupling}, \ref{fig:spacedelay2} and
\ref{fig:controldomains}). 

\subsubsection{Change of excitability}
\label{sec:changeYesWeCan}

In this section we clarify the influence of the control parameter $K$ on the
excitability of the system, within the approximation of effective parameters.
In particular the influence of $\tilde{\beta}$ and $\tilde{\gamma}$ is
investigated, since they change for all control schemes in opposite ways.
For $vv$ coupling the influence of $\tilde{\beta}$ dominates and for the
other coupling schemes the influence of $\tilde{\varepsilon}$ is decisive.

For the three coupling schemes $vv$, $vu$, and $uv$ the transformations from
the original system to that with effective parameters leave the fixed point
invariant, i.e. the fixed point values with respect to the transformed
variables $\tilde{u}$ and $\tilde{v}$ are the same as with respect to the
original one ($u$ and $v$). Since the fixed point depends only on $\beta$ and
$\gamma$, the condition that the fixed point remains the same forces the
effective parameters $\tilde{\beta}$ and $\tilde{\gamma}$ to change
simultaneously.  Thus they are not independent from each other. The parameters
$-\beta$ and $\frac{1}{\gamma}$ define the intersection with the $u$ axis and
the slope of the $v$-nullcline of the FHN system, respectively.  Thus, for a
smaller slope of the $v$-nullcline and unchanged fixed point, $\tilde{\gamma}$
has to be increased and $\tilde{\beta}$ decreased and vice versa. For the three
cases of $uv$, $vu$ and $vv$ the dependence between  $\tilde{\beta}$ and
$\tilde{\gamma}$ yields the condition 
\begin{equation}
  \label{eq:betaVonGamma}
  \tilde{\beta}(\tilde{\gamma})=\beta+(\gamma-\tilde{\gamma})v^*. 
\end{equation}
It is not intuitively clear how excitability changes. Increasing
$\tilde{\gamma}$ while respecting the invariant fixed point condition given by
Eq.(\ref{eq:betaVonGamma}), decreases $\tilde{\beta}$, which yields two
opposing effects on excitability.  In the following it is shown that under
the condition in Eq.~(\ref{eq:betaVonGamma}) the influence of $\tilde{\beta}$
dominates.

\begin{figure}[!tb] 
\includegraphics[width=0.35\textwidth]{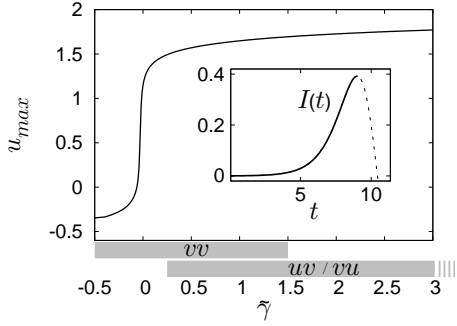}
\caption{
Maximum amplitude $u_{max}$ of activator $u$ after stimulation
with I(t)(inset: solid line).  System parameters: $\varepsilon$=$0.1$,
$\beta$=$\tilde{\beta}(K)$, $\gamma$=$\tilde{\gamma}(K)$. The bars below the
$x$-axis mark the ranges for the effective parameter $\tilde{\gamma}(K)$ for
$vv$ coupling and the two cross-coupling schemes $uv$ and $vu$. $K$ is varied in
the range $\left[-1,1\right]$.
} 
\label{fig:effpara}  
\end{figure}

To investigate the influence of simultaneously changing $\tilde{\beta}$ and
$\tilde{\gamma}$ with unchanged fixed point, numerical simulations of a single
FHN systems with effective parameters are performed under the condition in
Eq.~(\ref{eq:betaVonGamma}).  To excite the system the stimulation current
$I(t)$ (Inset in Fig.\ref{fig:effpara}) is added to the activator $u$.  The
response of the activator $u$ is determined in dependence of
$\tilde{\gamma}(K)$.  Note that these simulations are done without
feedback.  In order to deduce the influence of $\tilde{\beta}$ and
$\tilde{\gamma}$ upon excitability, $\tilde{\varepsilon}=\varepsilon=0.1$ is not
changed during the simulations.  Thus the chosen parameters are equivalent to
the effective parameter set of $vv$ coupling.  For the other system parameters
$\beta=0.85$ and $\gamma=0.5$ are chosen. 

In Fig.~\ref{fig:effpara} the maximum amplitude $u_{max}$ of the activator is
plotted versus $\tilde{\gamma}(K)$.  For $\tilde{\gamma}(K)<-0.05$ the
response of the system to the stimulation $I(t)$ has a small amplitude. For
$\tilde{\gamma}(K)>-0.05$ the amplitude of the system blows up in a very tiny
parameter range. After this \mbox{blow-up} the amplitude remains for all
$\tilde{\gamma}(K)>-0.05$ at about the same level.

This blow-up resembles the canard explosion of the FHN system near the Hopf
bifurcation.  The difference is that a canard transition usually characterizes
the fast blow-up of limit cycles, whereas in case of an excitable system a
limit cycle does not exist, but only its ghost is visible from the excited
trajectories. However, the result is that increasing $\tilde{\gamma}(K)$ with
simultaneously adjusting $\tilde{\beta}$ according to Eq.~(\ref{eq:betaVonGamma})
increases the response to a stimulus. We conclude that
excitability increases. Therefore it follows that the influence of decreasing
$\tilde{\beta}$ dominates.

Considering the ranges of $\tilde{\gamma}(K)$ that result for the different 
coupling schemes by varying $K$ in the interval $\left[-1,1\right]$ the impact
of each effective parameter on the coupling schemes can be determined.

For the $vv$ coupling scheme the $\tilde{\gamma}(K)$ is in the interval
$\left[-0.5,1.5\right]$ for $K \in \left[-1,1\right]$. The excitation of the
loop leads to the canard like blow-up close to $\tilde{\gamma}(K)=-0.05$, which
is equivalent to $K_{vv}=-0.55$.  Hence we estimate for $K_{vv}=-0.55$ a change
of excitability. This is in good agreement with the onset of successful control
for time-delayed feedback in the spatially extended system
(cf.~Fig.~\ref{fig:controldomains} d)).

For the two cross-coupling schemes $uv$ and $vu$ the change of
$\tilde{\varepsilon}$ dominates. $\tilde{\gamma}(K)$ lies in the interval
$\left[0.25,\infty\right[$.  The values of $\tilde{\gamma}$ for these two
schemes are beyond the canard-like transition, in the region of large
amplitudes. For the investigated values of $K$ the system does not undergo the
canard-like transition from large to small amplitudes and hence the influence
of $\tilde{\gamma}$ and $\tilde{\beta}$ on the excitability is small. Therefore
the influence of $\tilde{\varepsilon}$ is decisive.  To verify this assumption
we estimate the value of $K_{uv/vu}$ which is necessary to reach the
propagation boundary $\partial P$ by only considering $\tilde{\varepsilon}$.
The propagation boundary $\partial P$ for $\beta=0.85$ and $\gamma=0.5$ is
reached for $\varepsilon=0.1123$.  For larger $\varepsilon$ a stable pulse does
not exists.  Since the simulations were performed for $\varepsilon=0.1$, one
can compute the values of $K_{uv/vu}$ needed to move from $\varepsilon=0.1$ to
$\tilde{\varepsilon}=0.1123$, neglecting the influence of $\tilde{\beta}$ and
$\tilde{\gamma}$.  In the to cross-coupling schemes this yields
$K_{uv/vu}=\pm0.123$, which is in very good agreement with the results from the
simulations, where successful suppression is found for $\left| K_{uv/vu}
\right| >0.12$.  

Performing the same calculation for the $uu$ coupling assuming that also
in that case the influence of $\varepsilon$ dominates, $\tilde{\varepsilon}=0.1123$
is reached for $K=0.056$, whereas in the simulations already for $K>0.03$
successful control was observed, which is still of the correct order.

\subsubsection{Shift of propagation boundary}
\label{sec:pbs}

\begin{figure}[b]
\includegraphics[width=0.48\textwidth]{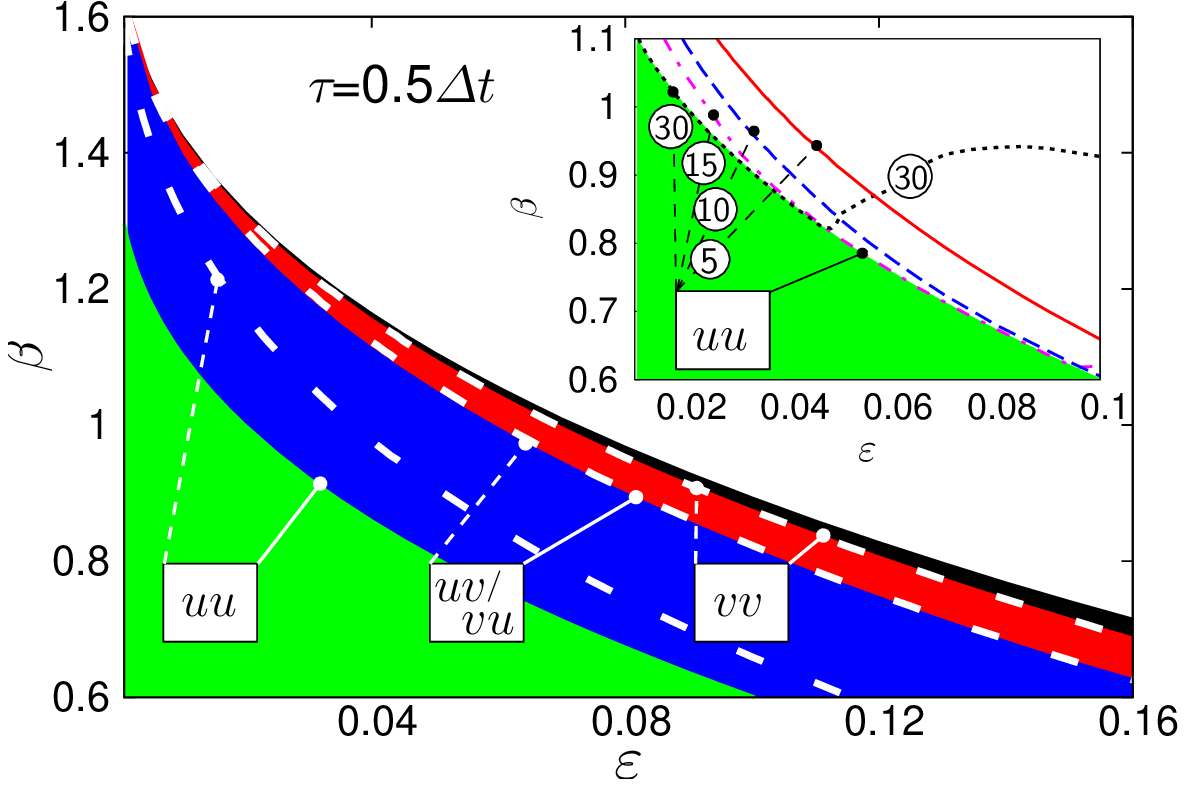}	
\caption{
Shift of propagation boundary $\partial P$: full lines: effective
parameters, dashed lines: with time-delayed feedback $\tau=0.5\Delta t$. Inset:
propagation boundary for $uu$ coupling, full line: with effective parameters,
dashed lines: with different delay times $\tau=5\Delta t$, $\tau$=$10\Delta t$,
$\tau$=15$\Delta t$, $\tau$=30$\Delta t$ ($\tau$ is given in units of the pulse
width $\Delta t$). Parameters: $\gamma=0.5$, $K=0.2$ for $uu$ and $uv$, $K=-0.2$
for $vu$ and $vv$.}
{\label{fig:pbs}} 
\end{figure}

Above we have introduced effective parameters by replacing the time-delayed
or space-shifted quantities in Eq.~(\ref{eq:TDAS}) and
Eq.~(\ref{eq:anisotropic_backward_coupling}) respectively, and have investigated the
influence on excitability for one set of system parameters ($\varepsilon=0.1$,
$\beta=0.85$, $\gamma=0.5$) and variable control parameter ($K$, $\delta$ and
$\tau$).  In this section the influence of the effective parameters on the
propagation boundary $\partial P$ in $(\varepsilon,\beta)$-parameter space (for
$\gamma=0.5$) is compared to the one obtained through simulations with
time-delayed feedback. Therefore $\left| K \right|=0.2$ and $\tau=0.5\Delta t$
is chosen, where $\Delta t$ is the temporal pulse width of the activator $u$.
The sign of the control parameter $K$ is chosen such that the
excitability is decreased, namely negative for $vv$ and $vu$ and positive for
$uu$ and $uv$.

The propagation boundaries are obtained in two different ways. Those with
time-delayed feedback are obtained by simulating the full equations and
performing a bisection method. Those propagation boundaries with effective
parameters are obtained by continuation of homoclinic orbits (stationary
pulse profiles) in the co-moving frame \cite{DAH08,SAN02b}. 

In Fig.~\ref{fig:pbs} the propagation boundaries $\partial P$ are shown. Those obtained
with effective parameters are the full lines separating different colors.  The white dashed
curves display the propagation boundaries for the full system
with time-delayed feedback. The boundaries are in good agreement for $vv$ coupling and
the cross-coupling schemes $uv$ and $vu$. For $uu$ coupling the two
propagation boundary differs from each other. The reason is that the effective
parameters are only exact for large $\tau$. 

The inset shows propagation boundaries of the $uu$ coupling. The full line
represents that obtained by effective parameters.  The different thin lines
represent the propagation boundaries obtained by simulating the full system
with different time delays.  In the range of small $\varepsilon$ for increasing
$\tau$ the displayed propagation boundaries of $uu$ coupling converge towards
the propagation boundary of the effective parameters.  The propagation boundary
with $\tau=30\Delta t$ perfectly fits the one obtained with effective
parameters for  $\varepsilon<0.04$.  However, for larger $\varepsilon$, this
boundary (dotted) diverges sharply from the one with effective parameters.
This is caused by the delay-induced bifurcation that occurs for large $K$ and
$\tau$, see Section~\ref{sec:single_system}.  Fig.~\ref{fig:pattern} (c) shows
the originating patterns that are already described in
Section~\ref{sec:delayInduOsci}.  After a local stimulation the time-delayed
feedback suppresses the emerging pulse, as predicted by the effective
parameters. But another effect occurs: the feedback is strong enough to create
a delay-induced oscillation.  This is due to the bistability of the local
system, where in addition to the stable fixed point a stable and an unstable
limit cycle are born in a saddle-node bifurcation. Each of the excitations can
spread for a small distance until it is again suppressed by the feedback.  Step
by step this pattern propagates and grows.

\section{Discussion and conclusion} 
\label{sec:discussion}

The increasing interest in both computational investigations of brain
functioning \cite{thisIssue} and control of complex dynamics
\cite{SCH07} has led us to combine methodologies and concepts from both
fields to investigate the pathogenesis and potential treatment
of brain disorders \cite{RUP99a}. This issue sets the context for our
studies. However, we expect that our results on controlling traveling pulses
can find also applications in other fields of biomedical engineering,  since
traveling pulses occur in many biological systems \cite{DAV90a,CAM93}.
Furthermore, in this study we consider models and methods of generic type, i.e., 
on the one side, we study pattern formation in reaction-diffusion systems
of activator-inhibitor type, namely the FitzHugh-Nagumo system. On the other
side stands a universal method of chaos control, that is, time-delayed feedback
(Pyragas control \cite{PYR92}), which is used to control the reaction-diffusion
patterns. In addition,  we consider nonlocal spatial coupling as a control
method. Nonlocal spatial coupling and time-delayed feedback have, as we have shown in
Sec.~\ref{sec:spcg}, a common mechanism that underlies the control of traveling
pulses.

The discussion will focus on two issues. Firstly,  the comparison of
time-delayed feedback and nonlocal spatial coupling, in particular the
predictive power of the effective parameters that can be introduced in the same
manner in both cases (Sec.~\ref{sec:dis1}). Secondly, the  time-delayed
feedback and nonlocal spatial coupling will be considered from a broader
perspective as augmented transmission capabilities in reaction-diffusion
systems with particular respect to the corresponding cortical structures
(Sec.~\ref{sec:dis2}).

\subsection{Time-delayed feedback and nonlocal spatial coupling}
\label{sec:dis1}

The control force $F(s)$ in Eq.~(\ref{eq:anisotropic_backward_coupling}),
i.\,e.,  the force in the anisotropic nonlocal type of coupling with backward
connections, can be directly compared to the control force $F(s)$  in
Eq.~(\ref{eq:TDAS}), which is time-delayed feedback (Pyragas control) applied
to each element in the active media locally. By going from 
Eq.~(\ref{eq:TDAS}) to  Eq.~(\ref{eq:anisotropic_backward_coupling}) nonlocal
connections are introduced simply by changing the position of the shift
operator from the first to the second argument of signal $s(x,t)$.  In other
words, Pyragas control is translated  from the temporal to the spatial domain,
as illustrated in Fig.~\ref{fig:coupling}.  This is compatible because the
pulse is stationary in the co-moving frame and thus the speed of the pulse
relates space to time scales.  If $\delta$ and $\tau$ are normalized to the
pulse width in the spatial and temporal domain, $\Delta x$ and $\Delta t$,
respectively, the common effect of nonlocal coupling and time-delayed feedback
on  traveling pulses is reflected in similar locations of the  control domains
in the control planes in Fig.~\ref{fig:spacedelay2}
and~\ref{fig:controldomains}.  

The analogy between  nonlocal coupling and time-delayed feedback, of
course, oversimplifies the situation.  However, the use of both types of
coupling has been proposed for control of spatio-temporal chaos in spatially
extended systems based on the idea of stabilization of unstable periodic
patterns embedded in spatio-temporal chaos \cite{LU96}. Using both types of
coupling simultaneously, it has been demonstrated through numerical analysis
that unstable roll patterns in a transversely extended three-level laser model
can be stabilized.  The motivation for this combined approach to control
unstable periodic patterns lies in the noninvasive character of both nonlocal
coupling and time-delayed feedback if the unstable periodic patterns is
approached. 

Our motivation to study nonlocal coupling and time-delayed feedback  is
somewhat different from that of chaos control, because we do not want to
stabilize unstable periodic patterns. We investigate the control of traveling
pulses by comparing nonlocal coupling with time-delayed feedback using both
types separately.   We suggest a method to predict the effect of the gain
factor $K$ on excitability by identifying the controlled system with the free
system with effective parameters, based on the idea that the effect of control
makes its main contribution upon the front dynamics.  Under this assumption, we
suggest to replace in both Eq.~(\ref{eq:anisotropic_backward_coupling}) and
Eq.~(\ref{eq:TDAS}) the shifted quantities  by their fixed point values,
i.\,e., $s(x,t-\tau)=s(x+\delta,t)=s^*$, where the signal $s$ was chosen to
be either $u$ or $v$ and $s^*$ is the corresponding fixed point value.

Despite its simplicity, this method yields accurate predictions for traveling
pulses invading the homogeneous steady state.  The resulting equations of
effective parameters, given in Tab.~\ref{tab:table}, describe by simple
algebraic relations how control changes excitability (Fig.~\ref{fig:phases}).
However, in which direction excitability changes with changing $K$ in the four
coupling schemes has to be investigated by evaluating the combined effect on
all three effective parameters $\tilde \varepsilon$, $\tilde \beta$ and
$\tilde \gamma$ (Fig.~\ref{fig:effpara}).  
 
The
noteworthy feature of our method is that the effective parameters describe a
 shift in the excitability of the local elements. 
By investigating time-delayed feedback in a single local element it is shown that
the onset of pulse suppression as well as the boundedness of the control
domains for large $K$ and $\tau$ can be explained by the local dynamics.
Furthermore, transferring the results of time-delayed feedback to the coupling type of
anisotropic nonlocal backward coupling one finds the same dependences.

\subsection{Reaction-diffusion with  augmented transmission as a hybrid model for the cortex}

\label{sec:dis2}

Control introduces augmented transmission capabilities in the
reaction-diffusion model  (Fig.~\ref{fig:controlScheme}).  The resulting
hybrid model in Eq.~(\ref{eq:main}) combines the two major signalling systems
in the brain, namely local coupling by diffusion, termed {\em volume transmission},
and {\em nonlocal coupling} in the spatial domain described by
Eqs.~(\ref{eq:isotropic_coupling}) and (\ref{eq:anisotropic_backward_coupling})
or in the temporal domain by Eq.~(\ref{eq:TDAS}).  The augmented transmission
capabilities are typical for synaptic transmission and neurovascular coupling.
For example, the change of sign in the gain parameter $K$ for pulse suppression
(Fig.~\ref{fig:isotropic_coupling}~b,c) is reminiscent of the  Mexican-hat type
functional and structural connectivity pattern in the cortex.  Also time-delays
of the order of seconds, that is, the order of  the width of the pulse profile
$\Delta t$ in spreading depolarizations, can occur in synaptic transmission if
metabotropic ion channels are involved, like metabotropic glutamate receptors,
which have increased open probabilities in the range of seconds after their
activation.  Moreover, typical latencies of this order result from the
neurovascular coupling. Therefore, the augmented transmission capabilities
represent internal neural circuity that is complementary to the volume
transmission introduced  in the original Hodgkin-Grafstein
Eqs.~(\ref{eq:singleSpecies})-(\ref{eq:cubic}) as a model for spreading
depolarizations.

The first hybrid model for spreading depolarizations that also combines the
two major signalling systems in the brain has been studied by Reggia and
Montgomery \cite{REG96}.  Potassium dynamics was modeled by a quadratic rate
function (cf.~Eq.~(\ref{eq:cubic})) and coupled to a neural network  that
mimics cortical dynamics and sensory map organization.  The key result of this
model is that at the leading edge of the potassium pulse, the elsewhere largely
uniform neural activity was replaced by a pattern of small, irregular patches
and lines of highly active elements, which could explain neurological symptoms
in migraine. However, there is no feedback of the network activity 
to the potassium reaction-diffusion pulse.
Therefore this hybrid models can not address the questions of the
controversially discussed Hodgkin-Grafstein mechanism \cite{STR05,HER05}
because the augmented transmission capabilities work only one way.  In a more
general context, such hybrid models are similar to creation
of spatio-temporal networks in addressable excitable media, which are studied
in the chemical Belousov-Zhabotinskii reaction \cite{TIN05}.

Volume transmission described by the Hodgkin-Grafstein mechanism was long
thought to be the main factor that causes the propagation in spreading
depolarizations \cite{STR05} although  synaptic transmission and gap junction
coupling were suggested to provide an alternative mechanism \cite{SHA01,HER05}.
Hybrid models can address  these controversial issue and may help to provide
insight into the spread and control of pathological pulses in the brain.  Our
emphasis is on understanding the role of internal cortical circuits that
provide augmented transmission capabilities and can prevent such events.
However, a long-term biomedical engineering therapeutic aim is also to design
strategies  that either support the internal cortical control or mimic its
behavior by external control loops and translate these methods into
applications.

\section*{Acknowledgements} 
This work was supported by DFG in the framework of Sfb 555. The authors
would like to thank Martin Gassel, Erik Glatt, Yuliya Dahlem, Steven Schiff,
Ken Showalter and Hugh Wilson for fruitful discussions.


\end{document}